\documentclass[twocolumn]{revtex4}
\usepackage{amssymb}
\usepackage{latexsym}
\usepackage{epsfig}
\usepackage{color}
\usepackage{float}
\usepackage{graphicx}
\usepackage{epstopdf}
\begin{document}
\title{Heat capacity of Holographic screen inspires MOND theory}
\author{M. Senay$^1$\footnote{mustafasenay86@hotmail.com}, M. Mohammadi Sabet$^2$\footnote{m.mohamadisabet@ilam.ac.ir}, H. Moradpour$^3$\footnote{Corresponding author: hn.moradpour@maragheh.ac.ir}}
\address{$^1$ Naval Academy, National Defence University, 34940, Istanbul, Turkey\\
$^2$ Ilam University, Ilam, Iran\\
$^3$ Research Institute for Astronomy and Astrophysics of Maragha
(RIAAM), University of Maragheh, P.O. Box 55136-553, Maragheh,
Iran}
\date{\today}
\begin{abstract}
It is argued that Planck mass may be considered as a candidate for
the mass content of degrees of freedom of holographic screen. In
addition, employing the Verlinde hypothesis on emergent gravity
and considering holographic screen degrees of freedom as a
$q$-deformed fermionic system, it is obtained that the heat
capacity per degree of freedom inspires the MOND interpolating
function. Moreover, the MOND acceleration is achieved as a
function of Planck acceleration. Both ultra-relativistic and
non-relativistic statistics are studied.
\end{abstract}
\pacs{....}
\maketitle

\section{Introduction}

The idea of Modified Newton dynamics (MOND theory) is simple,
Newtonian dynamics breaks down at low accelerations, and
therefore, a new constant is added to physics, usually called MOND
acceleration ($\equiv a_0$),
\cite{Milgrom1983a,Milgrom1983b,Milgrom1986c,rev1,beg,Milgrom:2019cle}.
Based on MOND theory, the second law of Newton ($F=ma$) is
modified as $F=ma\tilde{\mu}(a/a_0)$, where
$\tilde{\mu}\big(a/a_0(\equiv x)\big)$ is called the interpolating
function which should satisfy these conditions
\cite{Milgrom1983a,Milgrom1983b,Milgrom1986c,rev1}:

\begin{eqnarray}\label{mond}
&&\tilde{\mu}(x)\rightarrow\bigg\{^{1,\ x\gg1\ \big(\textmd{Newtonian limit}\big)}_{x,\ x\ll1\ \big(\textmd{deep MOND limit (DML)}\big)},\nonumber\\
&&\frac{d\ln\tilde{\mu}}{d\ln x}>-1.
\end{eqnarray}

\noindent In this manner, Poisson equation is also modified as \cite{rev1}

\begin{eqnarray}\label{mond1}
\nabla.\big[\tilde{\mu}(\frac{|\nabla\phi|}{a_0})\nabla\phi\big]=4\pi G\rho.
\end{eqnarray}

\noindent As a consequence, MOND theory has ability to explain the
rotational curves of spiral galaxies and their constant
luminosity, which are related to each other based on the Renzo
Sancisi's prediction \cite{rev1,Milgrom:2019cle}. The first
approximations, made by Milgorm, say $a_0$ is of order of
$10^{-8}\ cm\ s^{-2}$ \cite{Milgrom1983b}, and more studies claim
that $a_0=1.2 \times 10^{-8}\ cm\ s^{-2}$
\cite{beg,Milgrom:2019cle}.

Moreover, there is not any systematic way to find interpolating
function. The only things respected are Eq.~(\ref{mond}) and
compatibility with observations. On the other, debates on the
possible values of MOND acceleration is still ongoing
\cite{Rodrigues:2018duc,Chang:2018lab,Chan:2020gak,Marra:2020sts,Kroupa:2018kgv}.
Indeed, while Medium-richness galaxy groups can help us in testing
MOND at very low accelerations, it seems that this theory can not
solve the problem of mass in galaxy clusters and also rich galaxy
groups \cite{Milgrom:2019cle}. The latter may be considered as a
sign for this hypothesis that the mass content of sample is
important in getting true interpolating function. In summary, it
would mathematically be worthwhile to introduce a systematic way
for building interpolating function. In this regard, it is
addressed that the existence of a cutoff for acceleration ($a_0$)
may be due to the quantum statistics satisfied by the degrees of
freedom distributed on holographic screen \cite{Pazy2012}, or
even, may be considered as a sign of Debye gravity \cite{navia}.
On the other, it is also shown that MOND theory can be obtained if
generalized statistics is obeyed by the degrees of freedom of
holographic screen \cite{non}.

There are two distinct methods for investigating the statistics
and thermodynamics of intermediate states. First method is based
on Tsallis non-extensive statistics \cite{Tsallis1988}, and in
general, the generalized entropies \cite{pla}. In the second
method, the deformed quantum algebras are employed \cite{Arik1976}
leading to deformed thermostatistical functions. Recently, the
physical meaning of deformation parameters, generalized statistics
and their implications have been tried to be understood by
applying these deformed functions to various phenomena such as
condensed matter physics, solid state physics, nuclear physics,
gravitation and related topics
\cite{non,Neto2011,ob1,ob2,ob3,more1,more2,full,Shu2002,Zheng2012,Senay2018b,Kibaroglu2019}.

Here, adopting the Verlinde hypothesis on the emergence of gravity
and spacetime \cite{Verlinde2011}, and additionally, by relying on
the above arguments, we are looking at the holographic screen as a
system obeying $q$-statistics and study the predictions of this
view about MOND and address some related topics. In order to
achieve our aim, we begin with providing introductory notes on
Verlinde approach and $q$-statistics in the next two sections,
respectively. The effects of applying $q$-deformed fermion
statistics to holographic screen on MOND theory are also studied
in the forth section. The last section includes a summary.

\section{Verlinde gravity}

The essence of the Verlinde's idea is that gravity is no longer
fundamental and it appears as an entropic force (i.e. it emerges
as the tendency of systems to increase their entropy, in agreement
with the second law of thermodynamics) \cite{Verlinde2011}. We
consider a test particle with mass $m_t$. When this particle
approaches the holographic screen (it takes the distance $\Delta
x=\frac{\hbar}{m_tc}$ from the holographic screen
\cite{Verlinde2011}), the change of entropy of the holographic
screen is given by

\begin{equation}\label{1}
\Delta S=2\pi k_{B}\frac{m_tc}{\hbar}\Delta x,
\end{equation}

\noindent where $c$ is the speed of the light, and $k_{B}$ denotes
Boltzmann constant. There is a relation between temperature and
acceleration as \cite{Unruh1976}

\begin{equation}\label{2}
T=\frac{1}{2\pi}\frac{\hbar a}{k_{B}c}.
\end{equation}

\noindent For simplicity we consider energy units such that
$k_{B}=1$ throughout this paper. The particle feels the force
$F=T\frac{\Delta S}{\Delta x}=m_t a$. Moreover, since holographic
screen is a two-dimensional hypersurface with radius $R$ and area
$A=4\pi R^{2}$, the number of bits ($N$), distributed on it, is
calculated by

\begin{equation}\label{3}
N=\frac{Ac^{3}}{G\hbar}\equiv \frac{A}{l_p^2},
\end{equation}

\noindent where $G$ and $l_p\equiv\sqrt{\frac{G\hbar}{c^3}}$
denote the Newtonian gravitational constant and the Planck length,
respectively. The total energy of screen may be modelled by a one
dimensional Boltzmann gas \cite{Verlinde2011}

\begin{equation}\label{4}
E=\frac{1}{2}NT,
\end{equation}

\noindent which can also be found by using the mass ($M$) confined to it as

\begin{equation}\label{5}
E=Mc^{2}.
\end{equation}

\noindent Using the Eqs.~(\ref{2})-(\ref{5}), the Newtonian
gravity can finally be obtained as

\begin{equation}\label{6}
m_ta=T\frac{\Delta S}{\Delta x}\rightarrow a=G\frac{M}{R^{2}}.
\end{equation}

The classical relation~(\ref{4}), indeed a thermal energy, plays a
crucial role in getting the above result. Therefore, according to
Verlinde approach \cite{Verlinde2011}, gravitational effects are
related to the thermal excitations of the degrees of freedom of
holographic screen. Indeed, any modification to thermal
energy~(\ref{4}), corresponding to degrees of freedom, can affect
Eq.~(\ref{6}). Indeed, because the origin of holographic screen
and its degrees of freedom are not known, one may apply quantum
statistics or generalized statistics to its degrees of freedom to
obtain modifications to Eq.~(\ref{4}) and thus~(\ref{6})
\cite{Pazy2012,non,Neto2011,ob1,ob2,ob3}.

\section{$q$-deformed fermion gas \i n two d\i mens\i ons}

In this study, we propose a q-deformed fermionic system, which has
crucial application \cite{al}, and for example, in
Refs.~\cite{Algin2016a,full} some thermostatistical properties are
examined and the connection between fermionic $q$ deformation and
thermal effective mass of a quasi-particle are found out.

In terms of creation operator $a^{*}$, annihilation operator $a$,
and number operator $\hat{N}$, the symmetric $q$-deformed fermion
oscillators algebra is defined as \cite{full,more1,more2}

\begin{eqnarray}
&&aa^{*}+qa^{*}a=q^{-N},\ [\hat{N},a^{*}]=a^{*},\\
&& [\hat{N},a]=-a,\ a^{*}a=\hat{N},\ aa^{*}=1-\hat{N},\nonumber
\end{eqnarray}

\noindent where we have

\begin{eqnarray}
[x]=\frac{q^{x}-q^{-x}}{q-q^{-1}},
\end{eqnarray}

\noindent for the basic $q$-deformed quantum number. Here, $q$ is
real deformation parameter, and the Jackson derivative and mean
occupation number are as

\begin{eqnarray}
D_{x}^{q}f(x)=\frac{f(qx)-f(q^{-1}x)}{x(q-q^{-1})},
\end{eqnarray}

\noindent and

\begin{eqnarray}
f_{k,q}=\frac{1}{2\ln q}\ln\left[\frac{z^{-1}e^{\beta\varepsilon_{k}}+q}{z^{-1}e^{\beta\varepsilon_{k}}+q^{-1}}\right],
\end{eqnarray}

\noindent respectively, in which $\exp(\beta\mu)$ is used for the
fugacity $(z)$. In this statistics, if the degrees of freedom of
holographic screen meet the $\varepsilon=\alpha p^s$ relation,
then one can reach \cite{full}

\begin{eqnarray}\label{s6e1}
&&N=\frac{gA}{\lambda^2}h_\eta(z,q),\nonumber\\
&&U=\frac{\eta gA}{\lambda^2}Th_{\eta+1}(z,q),
\end{eqnarray}

\noindent where $\lambda=\frac{h\alpha^{1/s}}{\pi^{1/2}T^{1/s}}
\left[\frac{\Gamma(2)}{\Gamma(2/s+1)}\right]^{1/2} $ is the
generalized thermal wavelength \cite{full} and $\eta=\frac{2}{s}$,
for the number of degrees of freedom and their energy content in
two dimensions, respectively.

Defining $y=\beta\varepsilon$ and
$h_i(z,q)=\frac{1}{\Gamma(n)}\int_0^{\infty}\frac{y^{n-1}}{2\ln
q}\bigg[\ln\big(\frac{z^{-1}\exp(y)+q}{z^{-1}\exp(y)+q^{-1}}\big)\bigg]dy$,
which denotes the generalized Fermi integral, and by using
Eq.~(\ref{3}), one obtains

\begin{eqnarray}\label{s6e2}
&&h_\eta(z,q)=\frac{\lambda^2}{g\l_p^2},
\end{eqnarray}

\noindent as the equation which gives us fugacity $z$ and thus
chemical potential corresponding to the degrees of freedom of
holographic screen. The same assumption is also obtainable in
\cite{Pazy2012}.

In order to bring out the remarkable properties of the behavior of
the $q-$deformed fermion gas model at low temperatures, the generalized
Fermi integral can be calculated by using Sommerfeld expansion method.
So, we find

\begin{eqnarray}\label{s6e3}
h_{n}(z,q)=\frac{(lnz)^{n}}{\Gamma(n+1)}\left[1+n(n-1)\frac{\pi^{2}}{6}\gamma_{1}(q)\frac{1}{(lnz)^{2}}\right],
\end{eqnarray}

\noindent up to the first order, in which

\begin{eqnarray}\label{s6e4}
\gamma_{n}(q)=\frac{\int_{0}^{\infty}dy\frac{y^{n}}{2\ln
q}\ln\left[\frac{e^{y}+q}{e^{y}+q^{-1}}\right]}{\int_{0}^{\infty}dy\frac{y^{n}}{e^{y}+1}},
\end{eqnarray}

\noindent satisfying $\gamma_{n}(q)=1$ when $q=1$. At high
temperatures and independent of the value of $q$, Eq.~(\ref{s6e1})
leads to $U=\eta NT$ and the Boltzmann gas is recovered
\cite{full}.

The existence of minimum length (the Planck length denoted by
$l_p$) signals us to think the holographic screen degrees of
freedom as two-dimensional entities~\cite{Verlinde2011} meaning
that each degree of freedom carries energy $E_{HS}\simeq\Delta
E\approx\frac{\hbar}{t_p}=\frac{\hbar c}{l_p}$ combined with
assumption $E=mc^2$ to reach $m=m_p$, where $t_p$ and $m_p$ denote
the Planck time and the Planck mass, as the mass content of each
degree of freedom, respectively. In fact, if we accept $l_p$ as
the minimum length, then it will also be equal to the minimum wave
length ($\equiv\lambda_{min}$) meaning that $E=\frac{\hbar
c}{\lambda_{min}}=\frac{\hbar c}{l_p}$ is the energy of this wave.
Hence, one can look at $E$ as the energy content corresponding to
each degree of freedom with length $l_p$ (or equally, $E=E_{HS}$)
leading to the previous result on mass obtained by relying on
energy-time uncertainty relation.

In Ref.~\cite{Pazy2012}, by looking at the spacetime degrees of
freedom as a non-relativistic fermionic system with spin
degeneracy $g=1$ and Fermi energy $E_{F}$, author shows that the
heat capacity of the sample may evoke interpolating function and
thus MOND theory. Although author does not say anything about the
mass content of these degrees of freedom, we saw that Planck mass
is a probable candidate for it. This result can be combined with
the MOND acceleration of Pazy work ($\frac{12cE_{F}}{\pi\hbar}$)
to reach $\frac{24}{\sqrt{\pi}}a_p$ where $a_p$ denotes Planck
acceleration, as the MOND acceleration of Pazy. Therefore,
although his pioneering work opens a window towards a systematic
way to find interpolating function, its estimation of acceleration
is not satisfactory unless we do not confine ourselves to Planck
mass by considering masses greater than Planck mass.
Mathematically, the Pazy's work may be used to find interpolating
function, and by fitting its proposal to observations, one can
find out some estimations for $m$ and MOND acceleration.

\section{MOND theory as the heat capacity of holographic screen}

In fact, Eq.~(\ref{2}) claims that high temperature limit is
equivalent to the high acceleration limit, i.e. the territory of
Newtonian physics (or equally, $\tilde{\mu}=1$) compared to the
MOND physics (or equally, $\tilde{\mu}\neq1$)) which becomes
dominant at low accelerations. Therefore, a true theory (result)
should cover the Newtonian gravity at high temperatures
(accelerations).

At low temperatures, Eq.~(\ref{s6e1}) implies \cite{full}

\begin{eqnarray}\label{s6e6}
U=U_{0}+\frac{\pi^{2}}{6}\eta\gamma_{1}(q)N\frac{T^{2}}{\varepsilon_{F}},
\end{eqnarray}

\noindent where $U_{0}=\frac{\eta}{\eta+1}N\varepsilon_{F}$ is the
ground state energy, and thus the corresponding thermal energy is
obtained as

\begin{eqnarray}\label{s6e9}
U_{th}^{(q)}=\frac{\pi^{2}}{6}\eta\gamma_{1}(q)
N\frac{T^{2}}{\varepsilon_{F}},
\end{eqnarray}

\noindent where, $\varepsilon_{F}$ denotes the Fermi energy
defined as

\begin{eqnarray}\label{s6e7}
\varepsilon_{F}=\left[\frac{4\pi\hbar^{2}\alpha^{\eta}N}{gA}\right]^{1/\eta}=
\frac{\varepsilon_{F}^{0}}{g^{1/\eta}},
\end{eqnarray}

\noindent in which $\varepsilon_{F}^{0}$ denotes the Fermi energy
of the spin-less sample ($g=1$).

Eq.~(\ref{s6e9}) can be inserted in Eq.~(\ref{5}), to reach

\begin{eqnarray}\label{s6e11}
T^{2}=\frac{6Mc^{2}\varepsilon_{F}}{\pi^{2}\eta\gamma_{1}(q)N},
\end{eqnarray}

\noindent combined with Unruh relation~(\ref{2}), to get

\begin{eqnarray}\label{s6e12}
a^{2}=\frac{24Mc^{4}\varepsilon_{F}}{\hbar^{2}\eta\gamma_{1}(q)N}
\end{eqnarray}

\noindent for acceleration recovering the result of
Ref.~\cite{Pazy2012} at the limit of $\gamma_{1}(q=1)=1$. Now,
using~(\ref{3}) and~(\ref{s6e12}), one easily reaches at

\begin{eqnarray}\label{s6e14}
a^{2}=\left(\frac{6c\varepsilon_{F}}{\hbar\pi\eta\gamma_{1}(q)}\right)G\frac{M}{R^{2}}.
\end{eqnarray}

\noindent Comparing with the MOND equation
\cite{Milgrom1983a,Milgrom1983b, Milgrom1986c,rev1,Pazy2012} and
Eq.~(\ref{mond}), we easily see this limit is indeed DML
($\tilde{\mu}(x)=x$) that leads to

\begin{eqnarray}\label{s6e15}
&&a\left(\frac{a}{a_{0q}}\right)=G\frac{M}{R^{2}},\nonumber\\
&&a_{0q}=\frac{6c\varepsilon_{F}}{\hbar\pi\eta\gamma_{1}(q)},
\end{eqnarray}

\noindent and thus

\begin{eqnarray}\label{s6e16}
a_{0q}^{s=2}=\frac{6c\varepsilon_{F}}{\hbar\pi\gamma_{1}(q)},
\end{eqnarray}

\noindent for $\eta=1$ ($s=2$), and

\begin{eqnarray}\label{s6e17}
a_{0q}^{s=1}=\frac{3c\varepsilon_{F}}{\hbar\pi\gamma_{1}(q)},
\end{eqnarray}

\noindent for $\eta=2$ ($s=1$). In Ref.~\cite{Pazy2012},
considering $N_{Pazy}=\frac{A}{2l_p^2}$ instead of Eq.~(\ref{3})
and applying a non-relativistic Fermi-Dirac statistics ($s=2$) to
holographic screen, it has been obtained that
$a_{0p}=12\frac{c\varepsilon_{F}}{\pi\hbar}$ in agreement with the
$\gamma_{1}(q=1)=1$ limit of Eq.~(\ref{s6e16}) (apart from a
factor $2$ due to $N_{Pazy}=\frac{N}{2}$).

Now, based on Verlinde hypothesis \cite{Verlinde2011}, the force
experienced by the test particle with mass $m_t$ is calculable by
using

\begin{eqnarray}\label{f0}
F=T\frac{\Delta S}{\Delta x},
\end{eqnarray}

\noindent combined with Eq.~(\ref{2}),
$F=m_ta\tilde{\mu}(a/a_{0q})$, and $T\Delta S=\Delta E=C\Delta T$
to reach

\begin{eqnarray}\label{f1}
\tilde{\mu}(a/a_{0q})=\frac{C(a/a_{0q})}{2\pi}\frac{\Delta a}{a},
\end{eqnarray}

\noindent whenever, (using Eq.~(\ref{2})), $a$ and $a_{0q}$ are
defined as $\frac{2\pi cT}{\hbar}$ and $\frac{2\pi
cT_{0q}}{\hbar}$, respectively, and $C$ subsequently denotes the
heat capacity.

Bearing the first paragraph of this section in mind, and in order
to satisfy the Newtonian limit for which $C=\eta N$, one should
note that Eq.~(\ref{f1}) leads to $\tilde{\mu}=\frac{\eta
N}{2\pi}\frac{\Delta a}{a}$ that should become equal to $1$
(Newtonian limit for which $\tilde{\mu}=1$). This expectation
yields $\frac{\Delta a}{a}=\frac{2\pi}{\eta N}$ and thus

\begin{eqnarray}\label{f2}
\tilde{\mu}(a/a_{0q})=\frac{C(a/a_{0q})}{N\eta}=\bigg\{^{\frac{C(a/a_{0q})}{N},\
\eta=1}_{\frac{C(a/a_{0q})}{2N},\ \eta=2},
\end{eqnarray}

\noindent a result in agreement with Ref. \cite{Pazy2012} that
estimates $\tilde{\mu}\simeq\frac{C}{2N_{Pazy}}$ .

As we mentioned in the introduction, while Milgrom found $a_0$ is
of order of $10^{-8}\ cm\ s^{-2}$ \cite{Milgrom1983b}, the case of
$a_0=1.2 \times 10^{-8}\ cm\ s^{-2}$ is the most probable
\cite{beg,Milgrom:2019cle}, and indeed, the constancy (uniqueness)
of the value of MOND acceleration is still controversial
\cite{Rodrigues:2018duc,Chang:2018lab,Chan:2020gak,Marra:2020sts}.
On the other, gravity is the result of the existence of mass
(energy), and thus, the amount of mass is reflected in the power
of gravity. Hence, one may expect different values of $q$ for
different samples with distinct mass content.

\subsection*{Ultra-relativistic case ($\alpha=c$, $s=1$)}

\begin{figure}\label{mio_UR}
    \includegraphics[width=0.4\textwidth]{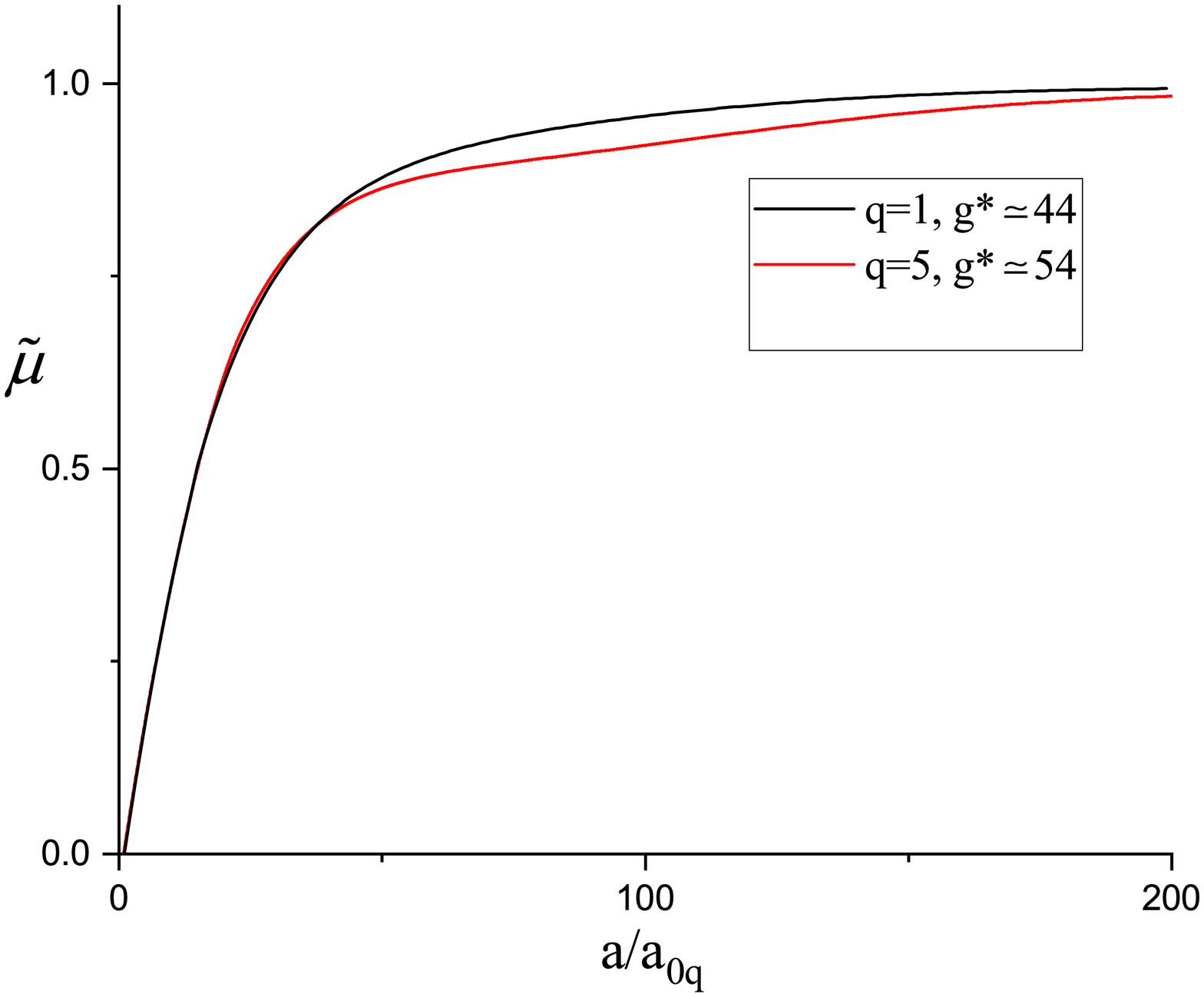}\\
    \includegraphics[width=0.4\textwidth]{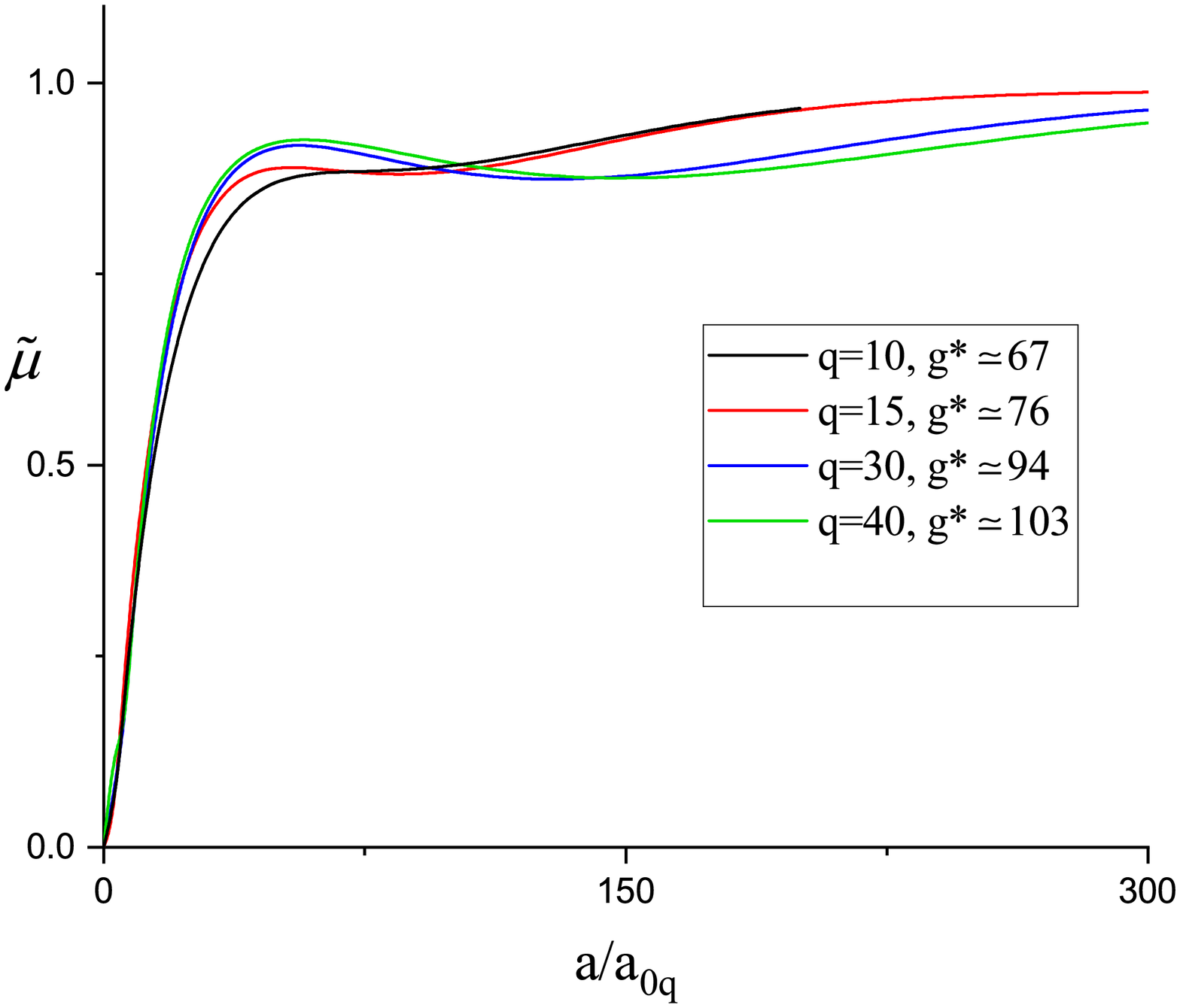}
    \caption{Interpolating function for ultra-relativistic formulation.}
\end{figure}

In order to continue, we consider the $\eta=2$ case, and plot
interpolating function for some values of $q$ and $g$ to show the
efficiency of idea. In this manner, from Eqs.~(\ref{3})
and~(\ref{s6e7}), we reach

\begin{eqnarray}
\!\!\varepsilon_{F}^0=\sqrt{4\pi}\frac{\hbar
c}{l_p}=\sqrt{\frac{4\pi\hbar c}{G}}\ c^2=\sqrt{4\pi}m_p\
c^2=\sqrt{4\pi}E_p,
\end{eqnarray}

\noindent where $m_p$ and $E_p$ denote the Planck mass and energy,
respectively. Now, we can write

\begin{eqnarray}\label{mondac1}
&&a_{0q}=\frac{3c}{\hbar\pi\gamma_1(q)g^{\frac{1}{2}}}\varepsilon_{F}^0=\frac{6}{\pi^{\frac{1}{2}}\gamma_1(q)g^{\frac{1}{2}}}a_p,\nonumber\\
&&T_{0q}=\frac{\hbar a_{0q}}{2\pi
c}=\frac{3}{\pi^{\frac{3}{2}}\gamma_1(q)g^{\frac{1}{2}}}T_p,
\end{eqnarray}

\noindent in which
$a_p=\sqrt{\frac{c^7}{G\hbar}}\simeq5.56\times10^{51}m/s^2$
denotes the Planck acceleration, $T_p=E_p$ is the Planck
temperature, and Eq.~(\ref{2}) has been employed to get the last
line. Thus, the value of MOND acceleration depends on the value of
$q$ and $g$.

Let us consider $a_{0q}=a_0=1.2 \times 10^{-8}\ cm\ s^{-2}$
\cite{beg,Milgrom:2019cle}, and use~(\ref{mondac1}) to obtain a
constraint on the values of $g(\equiv g^\star\times10^{122})$ and
$q$ as

\begin{eqnarray}\label{mondac1a}
\gamma_1(q)(g^\star)^{\frac{1}{2}}=\frac{6}{\pi^{\frac{1}{2}}a_0}a_p\simeq15.69.
\end{eqnarray}

\noindent For the general case including unknown MOND acceleration
$a_{0q}$ \cite{Rodrigues:2018duc}, the above equation takes the
form

\begin{eqnarray}\label{mondac1b}
\gamma_1(q)g^{\frac{1}{2}}=\frac{6}{\pi^{\frac{1}{2}}a_{0q}}a_p.
\end{eqnarray}

\noindent Simple calculations also lead to

\begin{eqnarray}\label{vel1}
\frac{GM}{r}=v^2\tilde{\mu}(\frac{v^2}{ra_{0q}}),
\end{eqnarray}

\noindent for the velocity $v$ of a particle moving under the
effect of source $M$.

If the value of MOND acceleration is constant, then
fitting~(\ref{vel1}) to observational data, and respecting
constraint~(\ref{mondac1a}), simultaneously, one can find proper
values of $g^\star$ and $q$. On the other, if the value of MOND
acceleration is not constant and known, then this approach has
three free parameters including $a_{0q}$, $g$ and $q$, found out
by fitting~(\ref{vel1}) to observations and also using
condition~(\ref{mondac1b}). Such analysis can give us worthwhile
info about $g$, $q$ and thus, the holographic screen nature.

\subsection*{Non-relativistic case}

\begin{figure}[h!]\label{mio_NR}
    \includegraphics[width=0.4\textwidth]{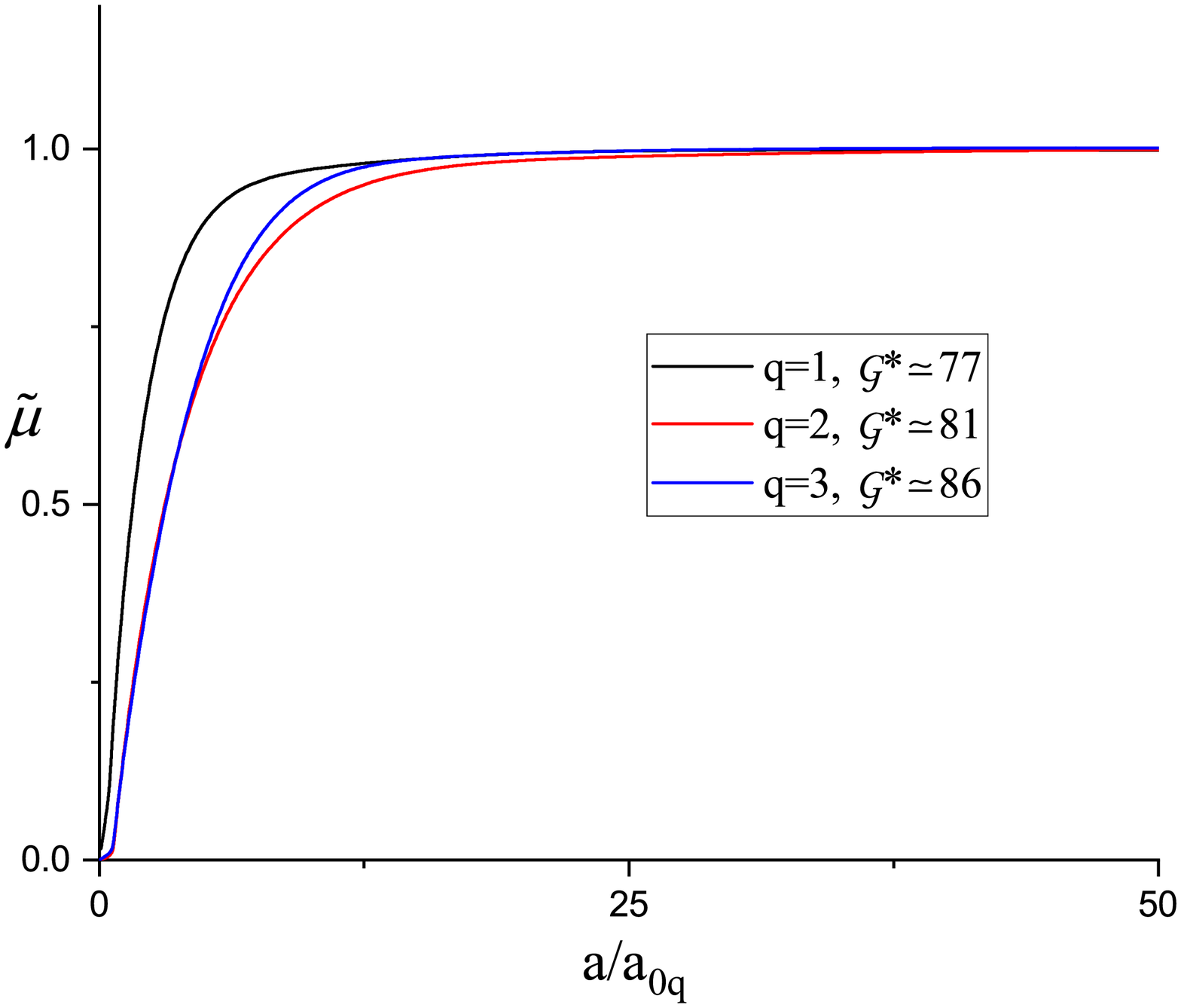}\\
    \includegraphics[width=0.4\textwidth]{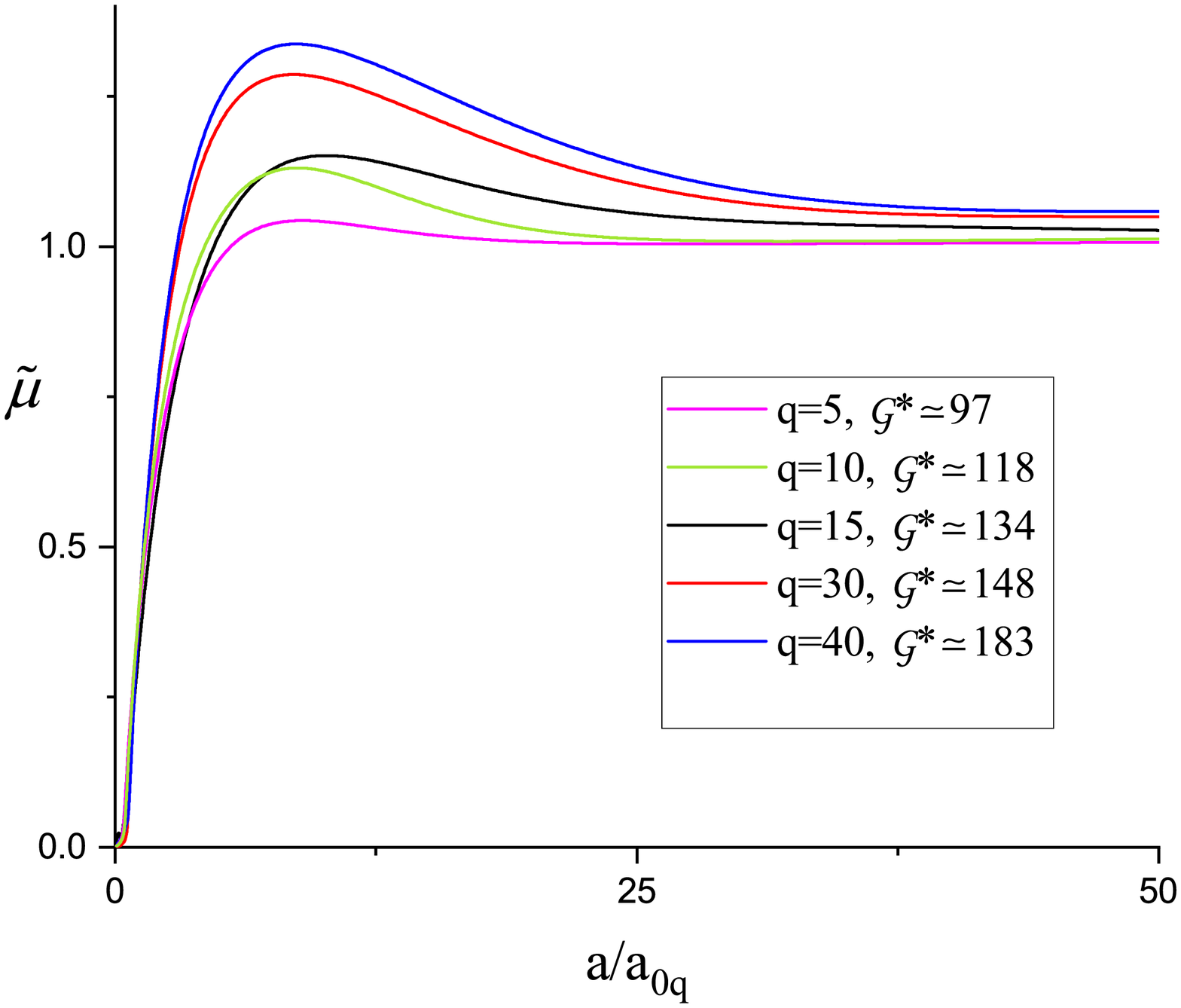}
    \caption{Interpolating function for non-relativistic formulation.}
\end{figure}

Here, $\alpha=\frac{1}{2m}$ and $s=2$ leading to
$\varepsilon_{F}^0=2\pi\frac{E_p^2}{mc^2}$, and thus, if we accept
our debate on the value of the mass of the degrees of freedom
distributed on the holographic screen, presented in the
introduction (or equally, $m=m_p$), then $\varepsilon_{F}^0=2\pi
E_p$. We can also see

\begin{eqnarray}\label{mondac2}
&&a_{0q}^{s=2}=\frac{12}{\gamma_{1}(q)g}(\frac{m_p}{m})a_p\Rightarrow a_{0q}^{s=2}=\frac{12}{\gamma_{1}(q)g}a_p\bigg|_{m=m_p},\nonumber\\
&&T_{0q}^{s=2}=\frac{\hbar a_{0q}^{s=2}}{2\pi
c}=\frac{6}{\pi\gamma_1(q)g}(\frac{m_p}{m})T_p\nonumber\\&&\Rightarrow
T_{0q}^{s=2}=\frac{6}{\pi\gamma_1(q)g}T_p\bigg|_{m=m_p},
\end{eqnarray}

\noindent leading to

\begin{eqnarray}\label{mondac2a}
&&\gamma_1(q)g=\frac{12}{a_0}(\frac{m_p}{m})a_p\Rightarrow \gamma_1(q)g=12\frac{a_p}{a_0}\bigg|_{m=m_p},\\
&&\gamma_1(q)g=\frac{12}{a_{0q}^{s=2}}(\frac{m_p}{m})a_p\Rightarrow
\gamma_1(q)g=12\frac{a_p}{a_{0q}^{s=2}}\bigg|_{m=m_p},\nonumber
\end{eqnarray}

\noindent as the counterparts of Eqs.~(\ref{mondac1a})
and~(\ref{mondac1b}), respectively. The counterpart of
Eq.~(\ref{vel1}) is also easily obtained by changing $a_{0q}$ with
$a_{0q}^{s=2}$. Therefore, in comparison with the
ultra-relativistic case, here, we have a new free parameter $m$,
if we do not accept the $m=m_p$ hypothesis.

\subsection*{Some primary solutions}

In Fig.~1 (Fig.~2), the Ultra-relativistic (non-relativistic)
interpolating function has been plotted for different values of
$q$ as well as $g^*$ ($\mathcal{G}^*=g\times10^{-61}$ in
non-relativistic case) by considering $a_q^0=1.2 \times 10^{-8}\
cm/s^2$. In the non-relativistic case, we have considered $m=m_p$.
It is clear that the overall behavior of interpolating function is
in complete agreement with MOND conditions~(\ref{mond}). Our
calculations show that $i$) the existence of $N\eta$ in the
denominator of~(\ref{f2}) guarantees the satisfaction of the
Newtonian limit ($\tilde{\mu}(x\gg1)\rightarrow1$), and moreover,
$ii$) in the ultra-relativistic case and for $q\gtrsim 10$, the
interpolating function has a maximum at a certain value of $x$ and
then tends to unity in such a way that the condition~(\ref{mond})
is satisfied. The maximum point occurs at $q\gtrsim 5$ for
non-relativistic interpolating function. It is clear that,
compared to ultra-relativistic case, the Newtonian condition is
satisfied at smaller values of $x$ meaning that the corresponding
interpolating function tends to $\tilde{\mu}=1$ faster than the
ultra-relativistic one.


\section{Conclusion}

Applying $q$-deformed fermionic statistics to the degrees of
freedom of holographic screen, and relying on the Verlinde
hypothesis, we found out that MOND interpolating function can be
understood as the heat capacity per degree of freedom of
holographic screen. Consequently, $i$) there are more free
parameters than Ref.~\cite{Pazy2012}, useful if one wants to fit
the theory with various observations, and $ii$) the classical
statistics (the Boltzmann gas) does only lead to Newtonian
gravity, because its heat capacity per degree of freedom is
constant.

On the other, it is also seen that the MOND acceleration is a
function of Planck acceleration, and additionally, Planck mass may
be accepted as a candidate for the mass content of each degree of
freedom of holographic screen. Therefore, in general, our approach
has two (three) free parameters in ultra-relativistic
(non-relativistic) case including $q$ and $g$ ($q$, $g$, and $m$),
if we do not limit ourselves to a special MOND acceleration
\cite{Rodrigues:2018duc,Chang:2018lab,Chan:2020gak,Marra:2020sts,Kroupa:2018kgv}
and $m$ (in non-relativistic case). In both cases, any
restrictions on the value of MOND acceleration (such as accepting
a value for it \cite{Milgrom:2019cle}) leads to eliminate one of
the parameters $q$ and $g$ with the help of Eqs.~(\ref{mondac1b})
and~(\ref{mondac2a}) in favor of the other.

It is finally worthwhile to mention that since a fermionic system
has been focused, $g$ should be even. Of course, this condition is
satisfied in our plots, but proper curves are obtained for very
huge amounts of $g$. Indeed, until we get a comprehensive (at
least a better) understanding of holographic screen, we can not
say anything about the allowed even values of $g$. On the other,
if we look at the results only as the mathematical functions that
meet the MOND requirements~(\ref{mond}), then we can even consider
odd values for $g$ to achieve a better fitting with observations.


\end{document}